\documentclass[prd,floats,preprint,superscriptaddress,eqsecnum,floatfix,nofootinbib,12pt]{revtex4}
\pdfoutput=1
\usepackage{amssymb,amsmath,amsthm,graphicx,ulem}
\usepackage{graphics, graphicx}
\usepackage{epsfig}
\usepackage{ulem}
\usepackage{color}
\usepackage{epstopdf}
\usepackage{latexsym}

\usepackage{braket}
\usepackage{tikz}
\usetikzlibrary{decorations.markings,decorations.pathmorphing}
\tikzset{->-/.style={decoration={markings,mark=at position #1 with {\arrow{>}}},postaction={decorate}}}
\tikzset{>=latex}

\newcommand{\D}{\ensuremath{\mathrm{d}}}

\def\be{\begin{equation}}
\def\ee{\end{equation}}
\def\beq{\begin{eqnarray}}
\def\eeq{\end{eqnarray}}
\def\p0{\phi_0}
\def\z0{\zeta_0}

\def\D1{D^{\ge 1}}

\newcommand{\ttle}[1]{}

\begin{document}

\title{\textbf{A Smooth Exit from Eternal Inflation?}}

\author{S. W. Hawking}
\affiliation{DAMTP, CMS, Wilberforce Road, CB3 0WA Cambridge, UK}
\author{Thomas Hertog}
\affiliation{Institute for Theoretical Physics, University of Leuven, 3001 Leuven, Belgium}

\bibliographystyle{unsrt}

\begin{abstract}

The usual theory of inflation breaks down in eternal inflation. We derive a dual description of eternal inflation in terms of a deformed Euclidean CFT located at the threshold of eternal inflation. The partition function gives the amplitude of different geometries of the threshold surface in the no-boundary state. Its local and global behavior in dual toy models shows that the amplitude is low for surfaces which are not nearly conformal to the round three-sphere and essentially zero for surfaces with negative curvature. Based on this we conjecture that the exit from eternal inflation does not produce an infinite fractal-like multiverse, but is finite and reasonably smooth.

\end{abstract}

\vskip.8in


\maketitle

\section{Introduction}
\label{intro}

Eternal inflation \cite{Vilenkin:1983xq} refers to the near de Sitter (dS) regime deep into the phase of inflation in which the quantum fluctuations in the energy density of the inflaton are large. In the usual account of eternal inflation the quantum diffusion dynamics of the fluctuations is modeled as stochastic effects around a classical slow roll background. Since the stochastic effects dominate the classical slow roll it is argued eternal inflation produces universes that are typically globally highly irregular, with exceedingly large or infinite constant density surfaces \cite{Linde:1996hg,Winitzki:2008zz,Creminelli:2008es,Hartle:2010vi}. 

However this account is questionable, because the dynamics of eternal inflation wipes out the separation into classical backgrounds and quantum fluctuations that is assumed. A proper treatment of eternal inflation must be based on quantum cosmology. In this paper we put forward a new quantum cosmological model of scalar field driven eternal inflation by using gauge-gravity duality \cite{Hull:1998vg,Balasubramanian2001,Strominger2001}. We define the Euclidean dual theory on the threshold surface of eternal inflation, which therefore describes the transition from the quantum realm of eternal inflation towards a classical universe, in line with the original vision behind inflation \cite{Brout:1977ix}. The subsequent evolution is assumed to be classical.

A reliable theory of eternal inflation is important to sharpen the predictions of slow roll inflation. This is because the physics of eternal inflation specifies initial conditions for classical cosmology. In particular a quantum model of eternal inflation specifies a prior over the so-called zero modes, or classical slow roll backgrounds, in the theory. This in turn determines its predictions for the precise spectral properties of CMB fluctuations on observable scales.

Our starting point remains the no-boundary quantum state of the universe \cite{Hartle1983}. This gives the ground state and is heavily biased towards universes with a low amount of inflation \cite{Hartle2008}. However we do not observe the entire universe. Instead our observations are limited to a small patch mostly along part of our past light cone. Probabilities for local observations in the no-boundary state are weighted by the volume of a surface $\Sigma_f$ of constant measured density, to account for the different possible locations of our past light cone \cite{Hartle2007}. This transforms the probability distribution for the amount of inflation and leads to the prediction that our universe emerged from a regime of eternal inflation \cite{Hartle2007,Hartle:2010dq}. Thus we must understand eternal inflation in order to understand the observational implications of the no-boundary wave function.

\begin{figure}[t]
\includegraphics[width=1.5in]{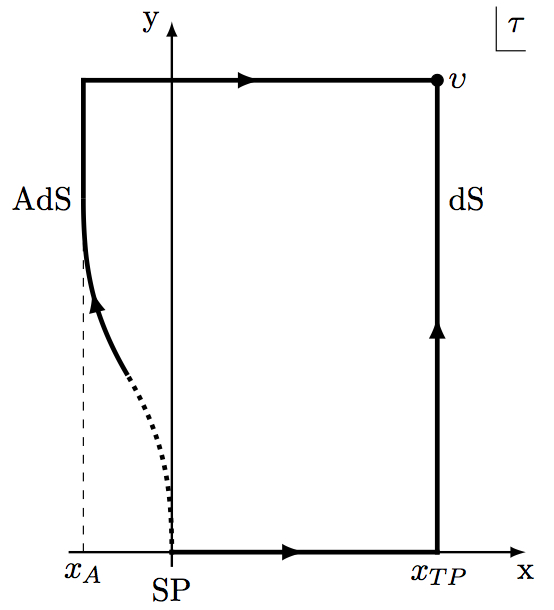} 
\caption{Two representations in the complex time-plane of the same no-boundary saddle point associated with an inflationary universe. The saddle point action includes an integral over time $\tau$ from the no-boundary origin or South Pole (SP) to its endpoint $\upsilon$ on $\Sigma_f$. Different contours for this give different geometric representations of the saddle point, each giving the same amplitude for the final real configuration $(h_{ij}(\vec x),\phi (\vec x))$ on $\Sigma_f$. The interior saddle point geometry along the nearly vertical contour going upwards from the SP consists of a regular, Euclidean, locally AdS domain wall with a complex scalar profile. Its regularized action specifies the tree-level probability in the no-boundary state of the associated inflationary, asymptotically de Sitter history. Euclidean AdS/CFT relates this to the partition function of a dual field theory yielding \eqref{dSCFT}. }  
\label{contour}
\end{figure}

However the standard saddle point approximation of the no-boundary wave function breaks down in eternal inflation. We therefore turn to gauge-gravity duality or dS/CFT \cite{Hull:1998vg,Balasubramanian2001,Strominger2001}, which gives an alternative form of the wave function evaluated on a surface $\Sigma_f$ in the large three-volume limit. In this, the wave function is specified in terms of the partition function of certain deformations of a Euclidean CFT defined directly on $\Sigma_f$. Euclidean AdS/CFT generalized to complex relevant deformations implies an approximate realisation of dS/CFT \cite{Maldacena2002,McFadden2009,Harlow2011,Maldacena2011,Hertog2011,Anninos2011}. 
This follows from the observation \cite{Hertog2011} that {\it all} no-boundary saddle points in low energy gravity theories with a positive scalar potential $V$ admit a geometric representation in which their weighting is fully specified by an interior, locally AdS, domain wall region governed by an effective negative scalar potential $-V$. We illustrate this in Fig. \ref{contour}. Quantum cosmology thus lends support to the view that Euclidean AdS/CFT and dS/CFT are two real domains of a single complexified theory \cite{Maldacena2002,Hull:1998vg,Dijkgraaf:2016lym,Bergshoeff:2007cg,Skenderis:2007sm,Hartle2012b}. In the large three-volume limit this has led to the following proposal for a holographic form of the semiclassical no-boundary wave function \cite{Hertog2011} in Einstein gravity,
\be
\Psi_{NB} [h_{ij}, \phi]= Z^{-1}_{QFT}[\tilde h_{ij},\tilde \alpha] \exp(iS_{st}[h_{ij}, \phi]/\hbar)   \ .
\label{dSCFT}
\ee
Here the sources $(\tilde h_{ij}, \tilde \alpha)$ are conformally related to the argument $(h_{ij}, \phi)$ of the wave function, $S_{st}$ are the usual surface terms, and $Z_{QFT}$ in this form of dS/CFT are partition functions of (complex) deformations of Euclidean AdS/CFT duals. The boundary metric $\tilde h_{ij}$ stands for background {\it and} fluctuations.

The holographic form \eqref{dSCFT} has led to a fruitful and promising application of holographic techniques to early universe cosmology (see e.g. \cite{Strominger2001b,McFadden2009,Bzowski:2012ih,Maldacena:2011nz,Garriga:2014fda,Afshordi:2016dvb}). No field theories have been identified that correspond to top-down models of realistic cosmologies where inflation transitions to a decelerating phase. However we find that many of the known AdS/CFT duals are ideally suited to study eternal inflation from a holographic viewpoint. This is because supergravity theories in $AdS_4$ typically contain scalars of mass $m^2=-2l^2_{AdS}$ with a negative potential for large $\phi$. In the context of \eqref{dSCFT} such scalars give rise to (slow roll) eternal inflation in the dS domain of the theory that is governed effectively by $-V$. In fact the Breitenlohner-Freedman bound in AdS corresponds precisely to the condition for eternal inflation in dS.

Here we use \eqref{dSCFT} to study eternal inflation holographically in toy-model cosmologies of this kind in which a single bulk scalar drives slow roll eternal inflation. We take the dual to be defined on a global constant density surface $\Sigma_f$ at the threshold (or exit) of the regime of scalar field driven eternal inflation. The bulk scalar driving inflation corresponds to a source $\tilde \alpha$ that turns on a low dimension scalar operator in the dual. Hence we use holography to excise the bulk regime of eternal inflation and replace this by field theory degrees of freedom on a kind of `end-of-the-world' brane. This is somewhat analogous to the holographic description of vacuum decay in AdS \cite{Maldacena:2010un}, although the interpretation here is different.

Conventional wisdom based on semiclassical gravity asserts that surfaces of constant scalar field in eternal inflation typically become highly irregular on the largest scales, developing a configuration of bubble-like regions with locally negative curvature. Holography provides a new perspective on this: The dependence of the partition function on the conformal geometry $h_{ij}$ of $\Sigma_f$ in the presence of a constant source $\tilde \alpha \neq 0$ specifies a {\it holographic measure} on the global structure of constant density surfaces in eternal inflation. We analyse various properties of this measure and find that the amplitude of surfaces with conformal structures far from the round one is exponentially small, in contrast with expectations based on semiclassical gravity. We also argue on general grounds that the amplitude is zero for all highly deformed conformal boundaries with a negative Yamabe invariant. This raises doubt about the widespread idea that eternal inflation produces a highly irregular universe with a mosaic structure of bubble like patches separated by inflationary domains.

\section{A Holographic Measure on Eternal Inflation}

\subsection{Setup}

For definiteness we start with the well known consistent truncation of M-theory on $AdS_4 \times S^7$ down to Einstein gravity coupled to a single scalar $\phi$ with potential
\begin{equation}\label{pot}
V(\phi)= -2-\cosh(\sqrt{2}\phi)\,,
\end{equation}
in units where $\Lambda=-3$ and hence $l^2_{AdS}=1$. The scalar has mass $m^2=-2$. Therefore in the large three-volume regime it behaves as
\begin{equation}\label{expansion}
\phi (\vec x, r)= \alpha (\vec x) e^{-r} + \beta (\vec x)  e^{-2r} + \cdots
\end{equation}
where $r$ is the overall radial coordinate in Euclidean AdS, with scale factor $e^r$. The Fefferman-Graham expansion implies that in terms of the variable $r$ the asymptotically (Lorentzian) dS domain of the theory is to be found along the vertical line $\tau = r+i\pi/2$ in the complex $\tau$-plane \cite{Hertog2011}. This is illustrated in Fig. \ref{contour} where $r$ changes from real to imaginary values along the horizontal branch of the AdS contour from $x_A$ to $x_{TP}$. This also means that in the dS domain the original potential \eqref{pot} acts as a positive effective potential
\begin{equation}\label{potdS}
\tilde V(\phi)= -V = 2+\cosh(\sqrt{2}\phi) \ .
\end{equation}
This is a potential for which the conditions for inflation and eternal inflation $\epsilon  \leq \tilde V$ hold, where $\epsilon \equiv \tilde V_{,\phi}^2/\tilde V^2$, for a reasonably broad range of field values around its minimum. This close connection between AdS supergravity truncations and eternal inflation in the framework of the no-boundary wave function \eqref{dSCFT} stems from the fact that the Breitenlohner-Freedman stability bound on the mass of scalars in $AdS$ corresponds precisely to the condition for eternal inflation in the de Sitter domain of the theory. 

Bulk solutions with $\phi \ll 1$ initially are at all times dominated by the cosmological constant $\Lambda$ and eternally inflate in a trivial manner. By contrast, solutions with $\phi \geq 1$ initially have a regime of scalar field driven eternal inflation, which eventually transitions into a $\Lambda$-dominated phase. The wave function \eqref{dSCFT} contains both classes of histories. We are mostly interested in the latter class and in particular in the amplitude of different (conformal) shapes of the constant scalar field transition surface\footnote{A realistic cosmology of course involves an intermediate radiation and matter dominated phase before the cosmological constant takes over. However since we are concerned with the structure of the universe at the exit from scalar field eternal inflation this toy-model setup suffices.} $\Sigma_f$ between these two regimes.

The variance of the semiclassical wave function of inhomogeneous fluctuation modes in the bulk is of order $\sim \tilde V/\epsilon$, evaluated at horizon crossing. In eternal inflation $\epsilon  \leq \tilde V$. Hence the fluctuation wave function spreads out and becomes broadly distributed \cite{Hartle:2010vi}. This is a manifestation of the fact that the universe's evolution, according to semiclassical gravity, is governed by the quantum diffusion dynamics of the fluctuations and their backreaction on the geometry rather than the classical slow roll \cite{Linde:1996hg,Winitzki:2008zz,Creminelli:2008es,Hartle:2010vi}. It is usually argued that the typical individual histories described by this wave function develop highly irregular constant density surfaces with a configuration of bubble-like regions with locally negative curvature. Below we revisit this from a holographic viewpoint.

We conclude this discussion of our setup with a few technical remarks. The argument $(h_{ij},\phi)$ of the wave function evaluated at $\upsilon$ in Fig. \ref{contour} is real. This means that in saddle points associated with inflationary universes, the scalar field must become real along the vertical dS line in the $\tau$-plane. The expansion \eqref{expansion} shows this requires its leading coefficient $\alpha$ to be imaginary, which in turn means that the scalar profile is complex along the entire interior AdS domain wall part of the saddle points. But the bulk scalar sources a deformation by an operator ${\cal O}$ of dimension one with coupling $\alpha$ in the dual ABJM theory. Hence the holographic measure in this model involves the AdS dual partition function on deformed three-spheres in the presence of an imaginary mass deformation $\alpha \equiv i\tilde \alpha$. We are primarily interested in the probability distribution over $\tilde h_{ij}$ for sufficiently large deformations $\alpha$, since these correspond to histories with a scalar field driven regime of eternal inflation. Finally whilst we formally define our dual on the exit surface $\Sigma_f$ from scalar field eternal inflation, at $\upsilon$ in Fig. \ref{contour}, we might as well take $\upsilon \rightarrow \infty$ because the classical, asymptotic $\Lambda$-phase amounts to an overall volume rescaling of the boundary surface which preserves the relative probabilities of different conformal bopundary geometries \cite{Hertog2011}.

\subsection{Local measure: perturbations around $S^3$}

We first recall the general behavior of partition functions for small perturbations away from the round $S^3$. Locally around the round sphere, the F--theorem and its extension to spin-2 deformations provide a general argument that the round sphere is a local minimum of the partition function. The F--theorem for three-dimensional CFTs  \cite{Jafferis:2010un,Klebanov2011} states that the free energy of a CFT on $S^3$ decreases along an RG flow triggered by a relevant deformation. A similar result was recently proved for metric perturbations of the conformal $S^3$ background \cite{Bobev:2017asb,Fischetti:2017sut}. The coupling of the energy-momentum tensor of the CFT to the curved background metric triggers a spin-2 deformation. The fact that the free energy is a local maximum for the round sphere is essentially equivalent to the positive definiteness of the stress tensor two-point function. Applied to the holographic no-boundary wave function \eqref{dSCFT} these results imply that the pure de Sitter history in the bulk is a local maximum of the holographic probability distribution, in contrast with expectations based on semiclassical bulk gravity in eternal inflation.

 \subsection{Global measure: squashed three-spheres}
 \label{squashed}
 
We now turn to large deformations. The dual of our bulk model is the ABJM SCFT. Hence to evaluate \eqref{dSCFT} we are faced with the problem of evaluating the partition function of supersymmetry breaking deformations of this theory. We do not attempt this here. Instead we first focus on a simplified model of this setup where we consider an $O(N)$ vector model. This is conjectured to be dual to higher-spin Vassiliev gravity in four dimensions \cite{Klebanov2002}. Higher-spin theories are very different from Einstein gravity. However, ample evidence indicates that the behavior of the free energy of vector models qualitatively captures that of duals to Einstein gravity when one restricts to scalar, vector or spin 2 deformations \cite{Hartnoll2005,Bobev:2016sap,Anninos2012}. This includes a remarkable qualitative agreement of the relation between the vev and the source for the particular scalar potential \eqref{pot} \cite{Conti:2017pqc}. We therefore view these vector models in this Section as dual toy models of eternal inflation and proceed to evaluate their partition functions for a specific class of large deformations. 
We return to Einstein gravity and a general argument in support of our conjecture below in Section \ref{general}.

Specifically we consider the $O(N)$ vector model on squashed deformations of the three-sphere,
\begin{align}
ds^2= \frac{r_0^2}{4} \left((\sigma_1)^2 +\frac{1}{1+A} (\sigma_2)^2 + \frac{1}{1+B}(\sigma_3)^2 \right) 
\label{eqn:metric}\;,
\end{align}
where $r_0$ is an overall scale and $\sigma_i$, with $i=1,2,3$, are the left-invariant one-forms of $SU(2)$. Note that the Ricci scalar $R(A,B)<0$ for large squashings \cite{Bobev:2016sap}. We further turn on a mass deformation ${\cal O}$ with coupling $\alpha$. This is a relevant deformation which in our dual $O(N)$ vector toy model induces a flow from the free to the critical $O(N)$ model. The coefficient $\alpha$ is imaginary in the dS domain of the wave function as discussed above. Hence we are led to evaluate the partition function, or free energy, of the critical $O(N)$ model as a function of the squashing parameters $A$ and $B$ and an imaginary mass deformation $\alpha \equiv \tilde{m}^2$. The key question of interest is whether or not the resulting holographic measure \eqref{dSCFT} favors large deformations as semiclassical gravity would lead one to believe. 

The deformed critical $O(N)$ model is obtained from a double trace deformation $f(\phi\cdot \phi)^2/(2N)$ of the free model with an additional source $\rho f \tilde{m}^2$ turned on for the single trace operator $\mathcal{O} \equiv (\phi \cdot \phi)$. By taking $f\rightarrow \infty$ the theory flows from its unstable UV fixed point, where the source has dimension one, to its critical fixed point with a source of dimension two \cite{Klebanov2002}. To see this we write the mass deformed free model partition function as
\begin{align}
	Z_{\rm free}[m^2]=\int \mathcal{D}\phi e^{-I_{\rm free}+\int d^3x\sqrt{g} m^2\mathcal{O}(x)} \ , \label{eqn:Zfree}
\end{align}
where $I_{\rm free}$ is the action of the free $O(N)$ model
\begin{align}
I_{\rm free}=\frac{1}{2}\int d^3x \sqrt{g} \left( \partial_{\mu} \phi_a \partial^{\mu}\phi^a +\frac{1}{8} R\phi_a \phi^a\right).
\end{align}
Here $\phi_a$ is an $N$-component field transforming as a vector under $O(N)$ rotations and $R$ is the Ricci scalar of the squashed boundary geometry. Introducing an auxiliary variable $\tilde{m}^2=\frac{m^2}{\rho f} +\frac{\mathcal{O}}{\rho}$ yields
\begin{align}
		Z_{\rm free}[m^2]=\int \mathcal{D}\phi \mathcal{D} \tilde{m}^2 e^{-I_{\rm free}+N \int d^3x\sqrt{g} \left[\rho f \tilde{m}^2\mathcal{O} -\frac{f}{2} \mathcal{O}^2-\frac{1}{2f}(m^2-\rho f \tilde{m}^2)^2\right] } \ , \label{eqn:ZfreeAux}
\end{align}
which can be written as
\begin{align}
	Z_{\rm free}[m^2]=\int \mathcal{D} \tilde{m}^2 e^{-\frac{N}{2f}\int d^3x\sqrt{g} (m^2-\rho f \tilde{m}^2)^2} Z_{\rm crit}[\tilde{m}^2]\ ,  \label{eqn:ZfreeifoZcrit}
\end{align}
with 
\begin{align}
	Z_{\rm crit}[\tilde{m}^2] =  \int \mathcal{D}\phi e^{-I_{\rm free}+N \int d^3x\sqrt{g}\left[\rho f \tilde{m}^2\mathcal{O} -\frac{f}{2} \mathcal{O}^2\right]} \ .
\end{align}
Inverting \eqref{eqn:ZfreeifoZcrit} gives $Z_{\rm crit}$ as a function of $Z_{\rm free}$:
\begin{align}
	Z_{\rm crit}[\tilde{m}^2] = e^{\frac{Nf\rho^2}{2}\int d^3x \sqrt{g} \tilde{m}^4}  \int \mathcal{D}m^2 e^{N\int d^3x\sqrt{g} \left(\frac{m^4}{2 f} -\rho \tilde{m}^2 m^2\right)}Z_{\rm free}[m^2] \ . \label{eqn:ZcritifoZfree}
\end{align}
The value of $\rho$ can be determined by comparing two point functions in the bulk with those in the boundary theory \cite{Anninos2012}. For the $O(N)$ model this implies $\rho=1$, which agrees with the transformation from critical to free in \cite{Anninos2013}.

We compute $Z_{\rm crit}$ for a single squashing $A \neq 0$ and $\tilde m^2 \neq 0$ by first calculating the partition function of the free mass deformed $O(N)$ vector model on a squashed sphere and then evaluate \eqref{eqn:ZcritifoZfree} in a large $N$ saddle point approximation\footnote{The generalization to double squashings $A,B \neq 0$ yields qualitatively similar results but requires extensive numerical work and is discussed in \cite{Conti:2017pqc}.}. Evaluating the Gaussian integral in \eqref{eqn:Zfree} amounts to computing the following determinant
\begin{align} 
 -\log Z_{\rm free}=F=\frac{N}{2}\log \left(\textrm{det}\left[ \frac{-\nabla^2 + m^2+\frac{R}{8}}{ \Lambda^2}\right]\right) \ , \label{eqn:LogZGeneral}
\end{align}
where $\Lambda$ is a cutoff that we use to regularize the UV divergences in this theory. The eigenvalues of the operator in \eqref{eqn:LogZGeneral} can be found in closed analytic form  \cite{Hu1973},
 \begin{align}
	\lambda_{n,q}= n^2+A (n-1-2q)^2-\frac1{4(1+A)} +m^2 \ , \quad q=0,1,\ldots , n-1,\ n=1,2,\ldots
\end{align}

To regularize the infinite sum in \eqref{eqn:LogZGeneral} we follow \cite{Anninos2012, Bobev:2016sap} and use a heat-kernel type regularization. Using a heat-kernel the sum over eigenvalues divides in a UV and an IR part. The latter converges and can readily be done numerically. By contrast the former contains all the divergences and should be treated with care. We regularize this numerically by verifying how the sum over high energy modes changes when we vary the energy cutoff. From a numerical fit we then deduce its non-divergent part which we add to the sum over the low energy modes to give the total renormalized free energy. The resulting determinant after heat-kernel regularization captures all modes with energies lower than the cutoff $\Lambda$. The contribution of modes with eigenvalues above the cutoff is exponentially small. For more details on this procedure we refer to \cite{Bobev:2016sap,Conti:2017pqc}. 

\begin{figure}[ht!]
\centering
    \includegraphics[width=0.45\textwidth]{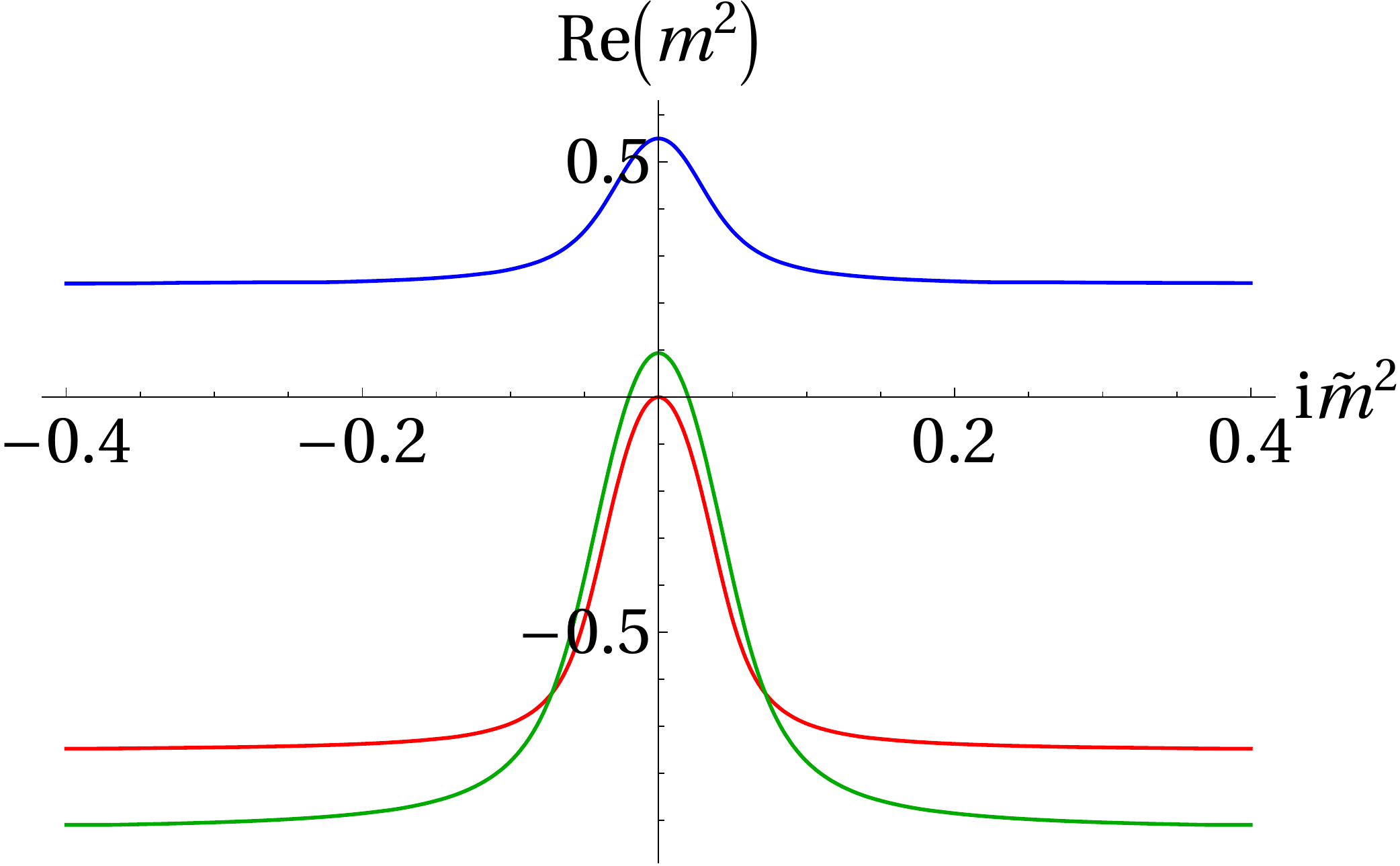}
    \includegraphics[width=0.45\textwidth]{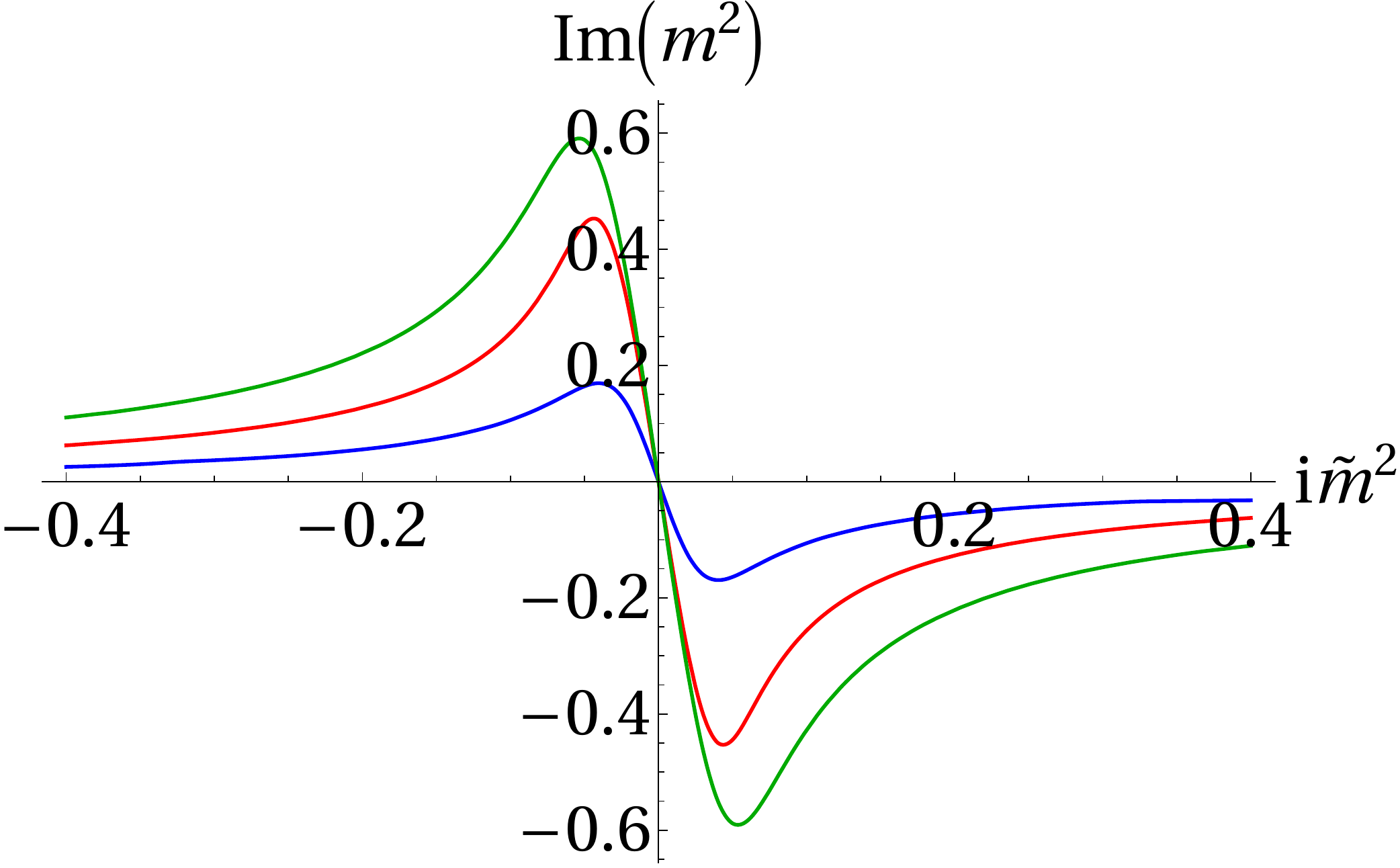}
\caption{The real and imaginary parts of the solutions $m^2$ of the saddle point equation \eqref{eqn:saddleSPN} are shown for three different values of a single squashing, i.e. $A=-0.8$ (blue), $A=0$ (red) and $A=2.06$ (green). For large $i\tilde{m}^2$ we have Re$(m^2) \rightarrow -R/8$.}\label{fig:saddles}
\end{figure}

To evaluate the holographic measure we must substitute our result for $Z_{\rm free}[A,m^2]$ in \eqref{eqn:ZcritifoZfree} and compute the integral in a large $N$ saddle point approximation. The factor outside the path integral in \eqref{eqn:ZcritifoZfree} diverges in the large $f$ limit. We cancel this by adding the appropriate counterterms. The saddle point equation then becomes
\begin{align}
\frac{2 \pi^2}{\sqrt{(1+A)(1+B)}} \left(\frac{m^2}{f}  - \tilde{m}^2\right) =- \frac{\partial \log Z_{\rm free}[m^2]}{\partial m^2} \ .\label{eqn:saddleSPN}
\end{align}

\begin{figure}[ht!]
\centering
    \includegraphics[width=0.65\textwidth]{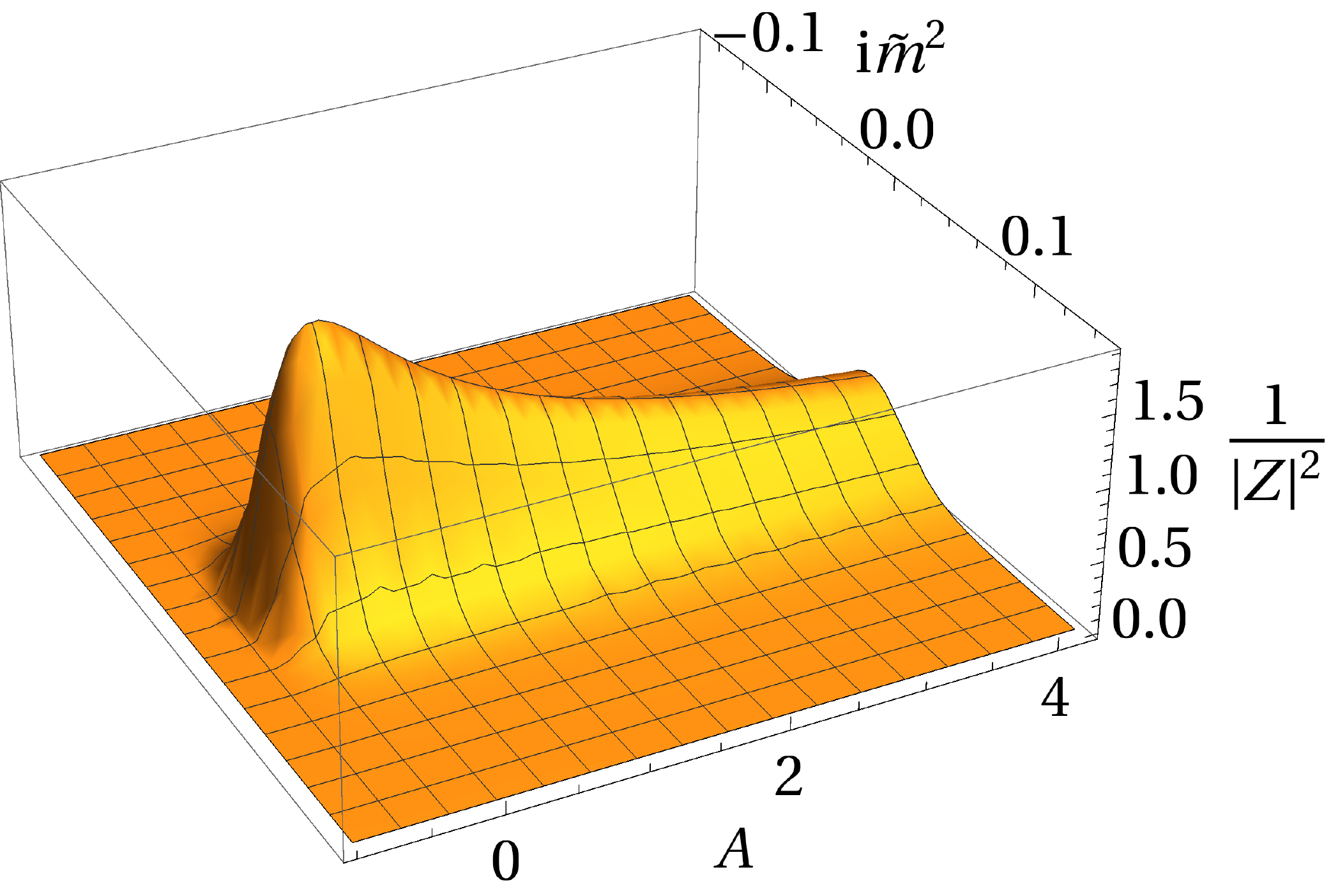}
    \caption{The holographic probability distribution in a dual toy model of eternal inflation as a function of the coupling of the mass deformation $\tilde{m}^2$ that is dual to the bulk scalar, and the squashing $A$ of the future boundary that parameterizes the amount of asymptotic anisotropy. The distribution is smooth and normalizable over the entire configuration space and suppresses strongly anisotropic future boundaries.}
  \label{fig:PvsBmsqSpN}
\end{figure}

We are interested in imaginary $\tilde{m}^2$ as discussed above. This means we need $Z_{\rm free}[A,m^2]$ for complex deformations $m^2$. Numerically inverting \eqref{eqn:saddleSPN} in the large $f$ limit we find a saddle point relation $m^2 (\tilde{m}^2)$. This is shown in Fig. \ref{fig:saddles}, where the real and imaginary parts of $m^2$ are plotted as a function of $i \tilde{m}^2$ for three different values of $A$. 

Notice that ${\rm Re}(m^2) \geq-R(A)/8$. This reflects the fact that the determinant \eqref{eqn:LogZGeneral}, which is a product over all eigenvalues of the operator $-\nabla^2 +m^2+R/8$, vanishes when the operator has a zero eigenvalue. Since the lowest eigenvalue of the Laplacian $\nabla^2$ is always zero, the first eigenvalue $\lambda_1$ of the operator in \eqref{eqn:LogZGeneral} is zero when $R/8 + m^2  =0$. In the region of configuration space where the operator has one or more negative eigenvalues the Gaussian integral \eqref{eqn:Zfree} does not converge, and \eqref{eqn:LogZGeneral} does not apply. This in turn means that the holographic measure $Z^{-1}_{crit}[A,\tilde m^2]$ is zero on such boundary configurations, as we now see.

\begin{figure}[ht!]
\centering
    \includegraphics[width=0.45\textwidth]{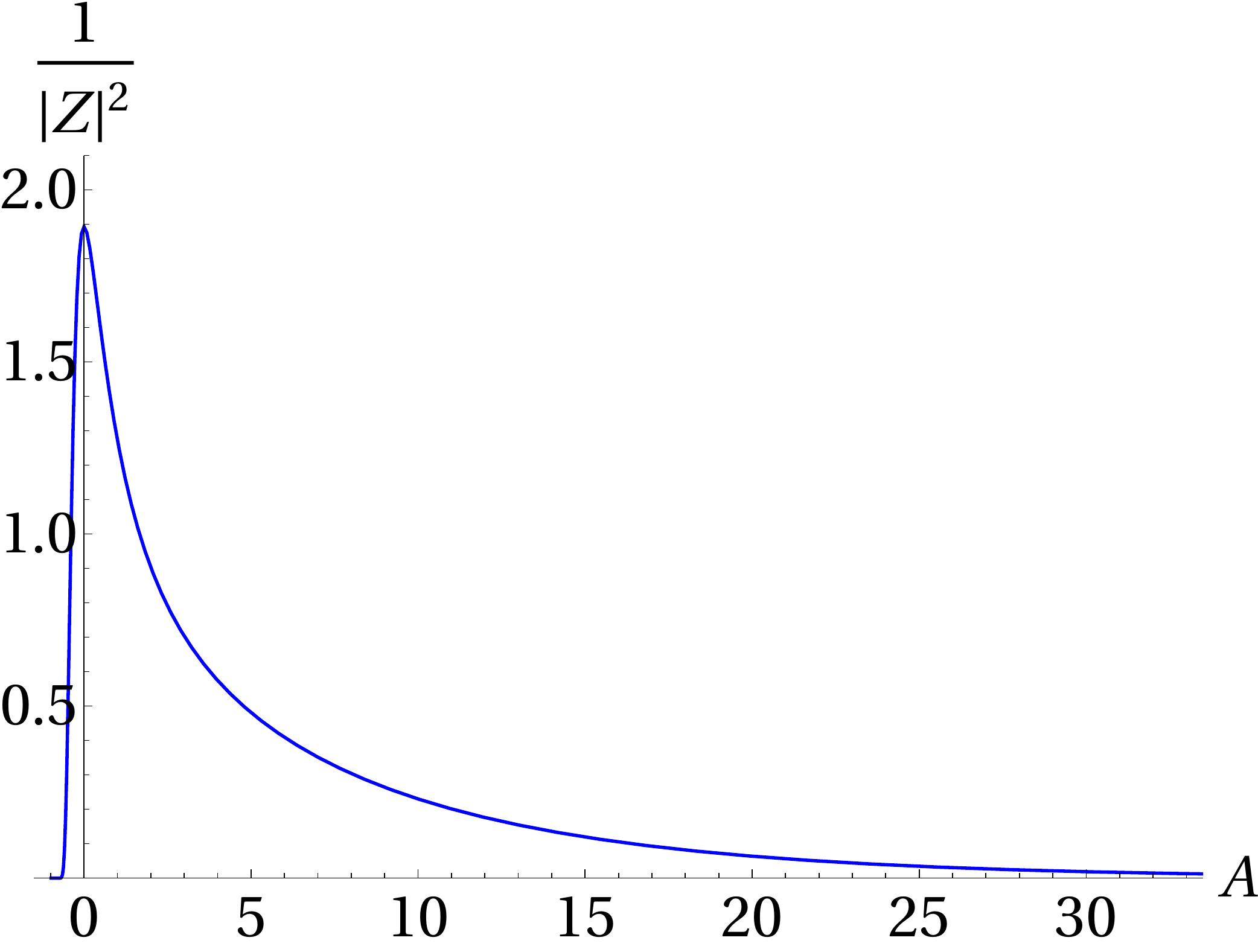}
    \includegraphics[width=0.45\textwidth]{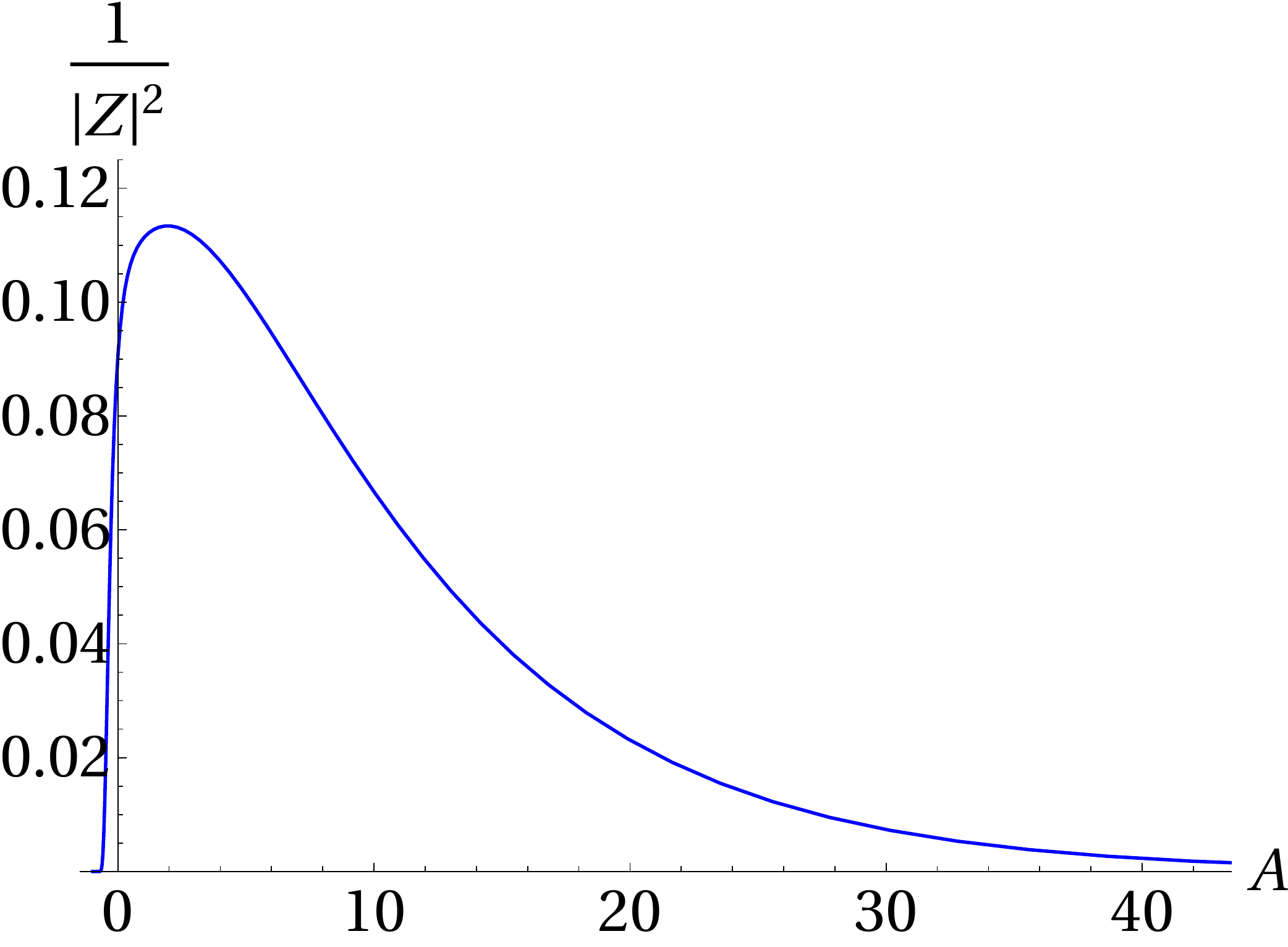}
\caption{Two slices of the probability distribution for $\tilde{m}^{2}=0.0$ (left) and $\tilde{m}^{2}=0.05i$ (right).}\label{fig:PvsBmsqSpNB}
\end{figure}

Inserting the relation $m^2 (\tilde{m}^2)$ in \eqref{eqn:ZcritifoZfree} yields the partition function $Z_{\rm crit}[A,\tilde m^2]$. We show the resulting two-dimensional holographic measure in Fig. \ref{fig:PvsBmsqSpN}. The distribution is well behaved and normalizable with a global maximum at zero squashing and zero deformation corresponding to the pure de Sitter history, in agreement with the F--theorem and its spin-2 extensions. When the scalar is turned on the local maximum shifts slightly towards positive values of $A$. However the total probability of highly deformed boundary geometries is exponentially small as anticipated\footnote{The distribution has an exponentially small tail in the region of configuration space where the Ricci scalar $R(A)$ is negative and $Z_{\rm free}$ diverges. We attribute this to our saddle point approximation of \eqref{eqn:ZcritifoZfree}.}. We illustrate this in Fig. \ref{fig:PvsBmsqSpNB} where we plot two one-dimensional slices of the distribution for two different values of $\tilde m^2$.  
 
\subsection{Global measure: general metric deformations}
\label{general}

It is beyond the current state-of-the-art to evaluate partition functions, be it of vector models or ABJM or duals to other models, for general large metric deformations. However, the above calculation implies a general argument suggesting that the amplitude of large deformations of the conformal boundary geometry is highly suppressed in the holographic measure both in higher-spin and in Einstein gravity. This is because the action of any dual CFT includes a conformal coupling term of the form $R\phi^2$. For geometries that are close to the round sphere this is positive and prevents the partition function from diverging. On the other hand the same argument suggests that the conformal coupling likely causes the partition function to diverge on boundary geometries that are far from the round conformal structure \cite{Witten1999}. These include in particular geometries with patches of negative curvature or, more accurately, a negative Yamabe invariant. 

The Yamabe invariant $Y(\tilde h)$ is a property of conformal classes. It is essentially the infimum of the total scalar curvature in the conformal class of $\tilde h$, normalized with respect to the overall volume. It is defined as
\be\label{YM}
Y(\tilde h) \equiv {\rm inf}_{\omega}\  {\cal I} (\omega^{1/4} \tilde h)
\ee
where the infimum is taken over conformal transformations $\omega(x)$ and ${\cal I} (\omega \tilde h)$ is the normalized average scalar curvature of $\omega^{1/4} \tilde h$,
\be
{\cal I} (\omega^{1/4} \tilde h) = \frac{\int_{M}\left(\omega^2 R(\tilde h)+8(\partial \omega)^2 \right)\sqrt{\tilde h} \ d^3_{\ } x}{\left(\int_{M}\omega^6 \sqrt{\tilde h} \ d^3_{\ } x \right)^{1/3}}
\ee
There always exists a conformal transformation $\omega (x)$ such that the metric $\tilde h'=\omega^{1/4} \tilde h$ has constant scalar curvature \cite{Schoen84}. The infimum defining $Y$ is obtained for this metric $\tilde h'$.

The Yamabe invariant is negative in conformal classes containing a metric of constant $R<0$. Since the lowest eigenvalue of the conformal Laplacian is negative on such backgrounds one expects that the partition function of a CFT does not converge, thereby strongly suppressing the amplitude of such conformal classes in the measure \eqref{dSCFT}. This is born out by the holographic measure specified by the partition function of the deformed $O(N)$ model on squashed spheres evaluated in Section \ref{squashed}. There the probabilities of large squashings for which $R<0$ are exponentially small, which can be traced in the calculation to the divergence of $Z_{free}$ on such backgrounds.

Conformal classes with negative $Y(\tilde h)$ precisely include the highly irregular constant density surfaces featuring in a semiclassical gravity analysis of eternal inflation. This general argument therefore suggests their amplitude will be low in a holographic measure. We interpret this as evidence against the idea that eternal inflation typically leads to a highly irregular universe with a mosaic structure of bubble like patches separated by inflationary domains\footnote{This resonates with \cite{Hartle:2010dq} where we argued that probabilities for local observations in eternal inflation can be obtained by coarse-graining over the large-scale fluctuations associated with eternal inflation, thereby effectively restoring smoothness. Our holographic analysis suggests that the dual description implements some of this coarse-graining automatically.}. Instead we conjecture that the exit from eternal inflation produces classical universes that are reasonably smooth on the largest scales.

\medskip

\section{Discussion}

We have used gauge-gravity duality to describe the quantum dynamics of scalar field driven eternal inflation in the no-boundary state in terms of a dual field theory defined on a global constant density surface at the exit from (scalar field) eternal inflation. Working with the semiclassical form \eqref{dSCFT} of dS/CFT the dual field theories involved are Euclidean AdS/CFT duals deformed by a complex low dimension scalar operator sourced by the bulk scalar driving eternal inflation. 

The inverse of the partition function specifies the amplitude of different shapes of the conformal boundary at the exit from scalar field eternal inflation. This yields a holographic measure on the global structure of such eternally inflating universes. We have computed this explicitly in a toy model consisting of a mass deformed interacting O(N) vector theory defined on squashed spheres. In this model we find that the amplitude is low for geometries far from the round conformal structure. Second, building on this result we have argued on general grounds that exit surfaces with significant patches of negative scalar curvature are strongly suppressed in a holographic measure in Einstein gravity too. Based on this we conjecture that eternal inflation produces universes that are relatively regular on the largest scales. This is radically different from the usual picture of eternal inflation arising from a semiclassical gravity treatment.

We have considered toy model cosmologies in which a scalar field driven regime of eternal inflation transitions directly to a $\Lambda$-dominated phase. The application of our ideas to more realistic cosmologies that include a decelerating phase requires further development of holographic cosmology (as is the case for all current applications of holographic techniques to early universe cosmology, e.g. \cite{Strominger2001b,McFadden2009,Bzowski:2012ih,Maldacena:2011nz,Garriga:2014fda,Afshordi:2016dvb}). It has been suggested that in realistic cosmologies, inflation corresponds to an IR fixed point of the dual theory \cite{Strominger2001b} in which case the partition function of the IR theory might specify the amplitude of exit surfaces. 

Our conjecture strengthens the intuition that holographic cosmology implies a significant reduction of the multiverse to a much more limited set of possible universes. This has important implications for anthropic reasoning. In a significantly constrained multiverse discrete parameters are determined by the theory. Anthropic arguments apply only to a subset of continuously varying parameters, such as the amount of slow roll inflation.

The dual Euclidean description of eternal inflation we put forward amounts to a significant departure from the original no-boundary idea. In our description, histories with a regime of eternal inflation have an inner boundary in the past, at the threshold for (scalar field) eternal inflation. The field theory on this inner boundary gives an approximate description of the transition from the quantum realm of eternal inflation, to a universe in the semiclassical domain. For simplicity we have assumed a sharp inner boundary, but of course one can imagine models where this is fuzzy. The detailed exit from eternal inflation is encoded in the coupling between the field theory degrees of freedom on the exit surface and the classical bulk dynamics.

\noindent{\bf Acknowledgments:} We thank Dio Anninos, Nikolay Bobev, Frederik Denef, Jim Hartle, Kostas Skenderis and Yannick Vreys for stimulating discussions over many years. SWH thanks the Institute for Theoretical Physics in Leuven for its hospitality. TH thanks Trinity College and the CTC in Cambridge for their hospitality. This work is supported in part by the ERC grant no. ERC-2013-CoG 616732 HoloQosmos.

\appendix

\bibliographystyle{thesis}
\bibliography{bibEEI.bib}

\end{document}